# Sonispace: a simulated-space interface for sound design and experimentation

Alexander Scarlatos

## ABSTRACT

The world of audio production and design has long been a difficult one to break into, requiring expertise and a working knowledge of the standard digital audio paradigms. This paper describes a novel interface that makes audio production and design more intuitive for novices, using sound-to-space relations that people have learned throughout daily life, such as the roles of barriers and distance in sound perception. The spatial interface for Sonispace allows users to quickly see the relationships between sound-emitting and sound-effecting objects, and to receive audio feedback as they make changes to the space. Algorithms were developed to resemble real-world sonic physics while being efficient enough to provide a user with immediate audio feedback. A prototype of the interface was tested by a group of participants, who confirmed that the software is accessible by novices and that the spatial interface is an engaging way of mixing audio.

## 1 INTRODUCTION

Audio software has long been dominated by linear tracks and stacked effects. While this paradigm provides a high level of control, it is optimized for a specific scenario: audio professionals constructing a polished, pre-assembled product. With the rise of more powerful digital technology, this standard has been expanded upon to include a variety of end users and contexts for audio. For example, programs like Ableton [1] have been developed so users can more easily construct a piece in a live setting. The context that Sonispace is designed for is the novice audio user or sound experimenter, looking to quickly and easily prototype a variety of soundscapes, specifically on a mobile device. This is achieved by presenting an intuitive, spatial environment for the user, where virtual objects represent physical ones in a digital sound space.

Similar environments have been developed, where object positions and qualities are direct controls for audio parameters. For example, Berthaut and Hatchet developed a 3D interface called Drile, where aspects of virtual objects, like size and color, are manipulated to alter musical qualities associated with those objects [2]. There are also mobile applications that are similar to Sonispace, like MelodyMorph [6], which is centered around musical nodes, and Singing Fingers [7], which is centered around drawing and manipulation of recorded audio. These types of interfaces are also referred to as "direct manipulation" systems, which have been shown to be significantly more usable by novices. Some desired features of these systems are the ability to "manipulate the object of interest directly and to generate multiple alternatives rapidly" and to use "simple metaphors, analogies, or models with a minimal set of concepts" [8]. Sonispace is unique because it uses the virtual space as a metaphor for a physical one, where sounds behave in a way similar to the real world. This is intended to familiarize users with the interaction concepts without needing prior experience of audio software.

## 2 INTERFACE OVERVIEW

The interface is centered around a simple metaphor: A user looks down into a room from the ceiling. In the room is a microphone ("receptor"), and the user hears whatever sound reaches it. The user can place objects that produce sound ("sound emitters") within the space, with those closest to the receptor sounding loudest. The user can also draw "walls" in the space, which will apply audio effects to the sound that reflects off or transmits through them. The interface is meant to act realistically, in that if a wall is far away from a sound emitter and the receptor, it will have little bearing on the change in sound. But if it is in their direct proximity, the effect will greater. The details of exactly how this relationship works are described in section 3 of this paper.

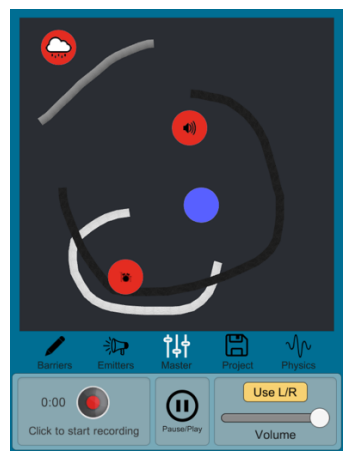

**Figure 1: Screenshot of the interface. The upper part of the screen contains the workable space and the bottom contains a toolbar. In the space you can see emitters (red nodes), the receptor (blue node) and walls (colored lines).**

For use in a practical audio mixing setting, we still need some concept of audio "tracks" and "effects". In Sonispace, each emitter carries an audio track that is either recorded in-app or imported from elsewhere. If the user wants multiple tracks playing at once, they just need multiple emitters in the space. Since the interface is designed for mobile devices, users will be able to easily record sounds in their daily lives and then work with those saved tracks. Sonispace also supports additional features such as looping and time syncing that give the user more control over the timing of audio tracks.

The walls carry the audio effects. When the user is drawing a wall, they have a certain "material" selected, which determines the audio effects associated with that wall. Each material has 2 effects: one for sound reflected off walls and one for sound transmitted through them. The user can change what these effects are and modify their parameters. By default, the spatial properties of the walls only change the dry/wet mix of the effects (volume of the original sound relative to the volume of the effected sound). However, the user may specify that other effect-specific aspects, such as delay time or phaser frequency, should change as well.

By tying the effect and audio track parameters to a space in these ways, there is a high degree of freedom, and many different ways to design a soundscape. Since the user can drastically change multiple parameters at once simply by dragging an object across the screen, this interface is ideal for rapid prototyping and experimenting with a variety of effect combinations.

## 3 IMPLEMENTATION

### 3.1 Physics-Based Modeling

In the real world, sound is the product of countless particles bumping into each other at different frequencies and intensities. The potentially high number of interactions is estimated using models, which convert object placements to audio track parameters. Some successful models are Wavefield Synthesis and High Order Ambisonics, which render audio for multi-speaker arrangements [3]. Sonispace uses its own model, which is designed to handle a constantly changing space.

Sonispace address two phenomena that people are familiar with when listening in daily life: reflection and diffraction. Reflection is how waves reverse direction upon hitting a barrier. If a receptor and emitter are on the same side of a barrier, and the receptor moves closer to the barrier, it will receive more reflected sound. Diffraction is how waves tend to bend around barriers and then normalize over a distance. If a receptor is directly behind a barrier across from a sound source, very little sound will be received. But as it moves back away from the barrier, it will receive more sound.

Instead of blocking a portion of a sound when it hits a barrier, Sonispace transmits modified sound through the barrier. The concept of diffraction is used to determine the amount of modified sound received relative to original sound received. This ratio is what the expected volume reduction would be in a realistic setting. The reason for this alteration is so users have more opportunities to effect their sound, and that blocking sound is a less desirable feature.

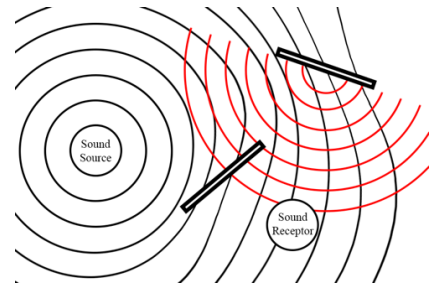

**Figure 2: In a realistic setting, sound will reflect off and "bend" around barriers.**

To represent these aspects of sound programmatically, a few easily implementable generalizations were made:
1. When an emitter and receptor are alongside a barrier:
   a. Moving the emitter or receptor closer to the barrier will increase the amount of reflected sound received.
2. When an emitter is across a barrier from a receptor:
   a. Moving the emitter or receptor closer to the barrier will decrease the amount of original sound.
   b. Moving either object closer to the barrier will also increase the amount transmitted sound.

The angle of incidence between an emitter and barrier also carries some impact:
1. Sound waves that hit the barrier at 90 degrees will have maximum reflection, and the amount of reflected sound decreases as the angle of incidence approaches 0.
2. Similarly, on the opposite side of the barrier, the most sound will be affected by transmission when the sound hits the barrier at a 90-degree angle, and the amount of sound affected by transmission will decrease as the angle of incidence approaches 0.

The same influences of angle apply equally to a receptor. When the angle of incidence decreases between a barrier and receptor, the sound reflected off or transmitted through that barrier will have less of an impact.

These generalizations are easily representable by providing each barrier with an "intensity" for each emitter, and the one receptor, in the space. This value represents how much the sound emitted from that node should be affected by the barrier. If any node moves further from the barrier, its intensity with that node will decrease. And as the angle between a barrier and node falls from 90 to 0 degrees, the intensity will decrease. To decide if the intensity will apply to reflected or transmitted sound, we only need to check to see if the receptor and emitter are on the same side of the barrier.

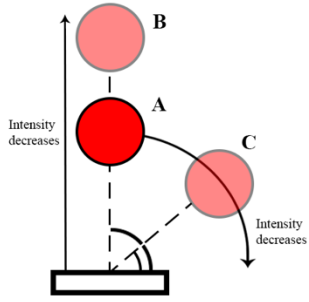

**Figure 3:** An emitter's relative position and angle to a barrier determines how much of its sound will be affected.

The walls in Sonispace are just chains of many small barriers. Because they are small, we can use a barrier's midpoint to measure its distance to a node. By summing the intensities of the barriers, we get the desired effect of a varying intensity while moving along a straight or curved surface, with the highest intensity given by being in proximity to the greatest surface area.

## 3.2 Parameterization Algorithm

The principle algorithm in Sonispace is the parameterization of the space, which takes the placements of the receptor $r$, all the barriers and emitters in the space, and outputs volume scalers for the audio tracks and effects. Audio effects are then applied directly to the audio tracks of the emitters, using these scalers (we will call them "mixes") to balance their volume, and the result is what the user hears. What we will end up with is a way to find the relative importance of each item in the space that runs in linear time.

The audio output of the program is the sum of, for each sound emitter $se$, a mix of its original data ($se.data$) and effected data (given by the $eff$ function).

$$Output = \sum_{se \,\in\, sound\ emitters} se.data * dMix_{se} + eff(se)$$

Each material is associated with 2 filters: one for reflection and one for transmission ($rFilter$ and $tFilter$), which are functions that will each apply a specific audio effect to a chunk of audio. The mix of these filter outputs is given by the relevance of the material $m$ in the space ($m.rMix_{se}$ and $m.tMix_{se}$), which is determined by the intensities of the barriers using that material. We will also need global mixes ($dMix_{se}$, $rMix_{se}$ and $tMix_{se}$) to determine how much total dry, reflected and transmitted sound is applied to the final mix.

$$eff(se) = \sum_{m \,\in\, materials} \begin{pmatrix} rMix_{se} * m.rMix_{se} * m.rFilter(se.data) \\ + tMix_{se} * m.tMix_{se} * m.tFilter(se.data) \end{pmatrix}$$

The first step of parameterization is calculating the intensities for each barrier in the space. The barrier's intensity relative to $se$, or its "emitter intensity" ($ei$) is multiplied with its intensity relative to $r$, or its "receptor intensity" ($ri$) to get its "total intensity" ($ti$). This way, a barrier will have a higher impact when the receptor and emitter are both in closer proximity to it.

$$b.ei = \frac{1}{dist(b,\ se)} * |\sin(angle(b,\ se))|$$

$$b.ri = \frac{1}{dist(b,\ r)} * |\sin(angle(b,\ r))|$$

$$b.ti = b.ei * b.ri$$

The function $dist$ gives the distance between $b$'s midpoint and a node, and the function $angle$ gives the angle between the line formed by the barrier's endpoints and the line between a node and the barrier's midpoint.

Once we have the barrier intensities, we can determine the relevance of each material. We keep two running values, reflected sum and transmitted sum ($m.rSum$ and $m.tSum$), for each material $m$. These are the sums of the total intensity of all barriers drawn with $m$, added to $m.rSum$ if the barrier is being used for reflection and added to $m.tSum$ if being used for transmission. We also keep track of the sum of these values across all materials ($rTotal$ and $tTotal$).

> for $b$ in barriers:
>     if $r$ and $se$ on same side of $b$:
>         $b.m.rSum_{se}$ += $b.ti$
>         $rTotal_{se}$ += $b.ti$
>     else if $r$ and $se$ on opposite sides of $b$:
>         $b.m.tSum_{se}$ += $b.ti$
>         $tTotal_{se}$ += $b.ti$

Once all barriers have been accounted for, we can obtain a reflected and transmitted mix for each material ($m.rMix_{se}$ and $m.tMix_{se}$). These keep track of how much each material relatively contributes to the total reflected and transmitted intensities of the space, and are used as the volume scalers for their audio effects (as shown in the $eff$ equation earlier).

$$m.rMix_{se} = \frac{m.rSum_{se}}{rTotal_{se}} \qquad m.tMix_{se} = \frac{m.tSum_{se}}{tTotal_{se}}$$

Once we have accounted for every material, we can calculate our global mixes.

$$rMix_{se} = \left(\frac{rTotal_{se}}{rTotal_{se} + tTotal_{se}}\right) * \left(\frac{rTotal_{se}}{rTotal_{se} + c}\right)$$

We calculate $tMix_{se}$ the same way but by swapping $rTotal_{se}$ and $tTotal_{se}$.

The first term in this equation keeps the ratio between reflected and transmitted mixes in check. Since we are keeping track of the mixes of materials relative to the total *reflected* and *transmitted* intensities, we can weight them properly by multiplying them with the global $rMix_{se}$ and $tMix_{se}$ (as shown in the $eff$ equation earlier).

The second term allows the mixes to grow relative to an adjustable constant $c$, so that a larger total intensity (more barriers in the space) will yield a higher mix of effected sound.

$$dMix_{se} = \frac{1}{dist(se,\ r)} * \Big(1 - (rMix_{se} + tMix_{se}) * d\Big)$$

The dry mix, *dMix*, increases as the distance between *se* and *r* decreases, and can also be offset by *rMix* and *tMix*, relative to an adjustable constant *d*. This constant lets the user decide how much they want to hear the original sound when the effects begin to kick in.

## 4   USABILITY EXPERIMENT

A preliminary usability experiment was conducted, where a group of participants had 45 minutes to create a soundscape using provided audio samples. At the completion of their soundscape, each user was requested to submit a recording of it, as well as answer a questionnaire. They were asked about their prior level of expertise with audio software, if they enjoyed Sonispace, and what aspects they liked most and least about the experience.

Out of the 23 responses received, 12 participants had little to no expertise with audio software, 9 had some expertise, and a negligible sample of 2 had significant expertise. About 78% of participants said they enjoyed using the software, and about 65% said they would use the software if it was polished. Interestingly, less experienced users in the group were more likely to enjoy the software, with 83% of low expertise users enjoying the software, 78% of some expertise users enjoying the software, and 50% of high expertise users enjoying the software. This indicates that Sonispace succeeded in its goal of being accessible to novices, but that it would need modifications in order to appeal to professionals.

When participants were asked about what they enjoyed most during the session, the most common responses were 1) being able to drag sound nodes, 2) being able to draw the effects, and 3) the simplicity of the interface. When asked about what needed work, the most common responses were 1) a lack of instructions in the app, 2) unintuitive controls, and 3) a lack of features. These results indicate that the fundamental idea of the interface was successful, but the implementation would need work to be more user friendly.

## 5   CONCLUSIONS AND FUTURE WORK

Sonispace is a unique interface that allows users to quickly immerse themselves in a world of sound, with the ability to record on the fly and experiment with a dynamic soundscape. The core algorithm provides a high degree of freedom, and the provided tools allow users to construct a project live and visualize its organization. A preliminary usability study demonstrated that the spatial metaphor of the software was a success, especially for users who are new to audio software.

One of the original inspirations for Sonispace was to use it in a social media context, where users could create soundscapes and share them with peers and collaborate on projects virtually. This system would be similar to the sharing and collaboration system used by Scratch [5]. The visual art medium has also seen mainstream success through apps that enable people people to share images, notably Instagram [4]. Instagram's simple interface on a mobile device allows any individual to try their hand at photography and photo effecting, with the social gratification of using their work to interact with an audience. Since mobile devices are also equipped with microphones, there is a high potential for a similar app in the audio realm, one that could provide the ability to quickly improve the quality of your recording and share with friends. Another phenomenon that has come into the mainstream recently is live streaming, as shown by the growing popularity of Facebook Live. The paradigm of live content creation could fit with Sonispace very easily, and a future effort would be to allow users to feed live audio into emitters and broadcast to an audience of followers.

## A   SUPPLEMENTAL VIDEO

Software Overview with Live Demo: https://youtu.be/WWau5UCgupo

## ACKNOWLEDGMENTS

This work would not have been possible without the guidance and help of some very supportive faculty at Stony Brook University. Dr. Margaret Schedel advised me during the undergraduate portion of the work, helping me define the project from a sound design perspective, pointing me in the right direction on how to simulate sound in a virtual space, and brainstorming different ideas for the project with me. Dr. Roy Shilkrot advised me during the graduate portion of the work, guiding me on how to conduct research for the project, helping me define a research context for it, pushing me to submit the work to conferences, and reviewing the paper. I would also like to thank Dr. Lori Scarlatos for helping me organize the usability experiment.